\documentclass[twocolumn,showpacs,preprintnumbers,amsmath,amssymb]{revtex4}
\usepackage{graphicx}
\usepackage{dcolumn}
\usepackage{bm}
\usepackage{lipsum}
\begin{document}

\title[Study of elastic and inelastic scattering of $^7$Be + $^{12}$C]{Study of elastic and inelastic scattering of $^7$Be + $^{12}$C at 35 MeV}

\author{K. Kundalia$^1$}
\email{kkabita@jcbose.ac.in}
\author{D. Gupta$^1$}%
\email{dhruba@jcbose.ac.in}
\author{Sk M. Ali$^1$}
\author{Swapan K Saha$^1$}
\altaffiliation {Former faculty}
\author{O. Tengblad$^2$}
\author{J.D.~Ovejas$^2$}
\author{A. Perea$^2$}
\author{I. Martel$^3$}		
\author{J. Cederkall$^4$}
\author{J. Park$^4$}
\altaffiliation{Present Address: Center for Exotic Nuclear Studies, Institute for Basic Science, 34126 Daejeon, South Korea}
\author{S. Szwec$^{5,6}$}
 \author{A.~M.~Moro$^{7,8}$}

\affiliation{$^1$Department of Physics, Bose Institute, 93/1 APC Road, Kolkata 700009, India}
\affiliation{$^2$Instituto de Estructura de la Materia $-$ CSIC, Serrano 113 bis, ES-28006 Madrid, Spain}
\affiliation{$^3$University of Huelva, Av. Fuerzas Armadas s/n. Campus ``El Carmen'', 21007, Huelva, Spain}
\affiliation{$^4$Department of Physics, Lund University, Box 118, SE-221 00 Lund, Sweden}
\affiliation{$^5$Accelerator Laboratory, Department of Physics, University of Jyv$\ddot a$skyl$\ddot a$, FI-40014 Jyv$\ddot a$skyl$\ddot a$, Finland}
\affiliation{$^6$Helsinki Institute of Physics, University of Helsinki, FIN-00014 Helsinki, Finland}
 \affiliation{$^7$Departamento de F\'{\i}sica At\'omica, Molecular y Nuclear, Facultad de F\'{\i}sica, Universidad de Sevilla, Apartado 1065, E-41080 Sevilla, Spain}
 \affiliation{$^8$Instituto Interuniversitario Carlos I de F\'isica Te\'orica y Computacional (iC1), Apdo.~1065, E-41080 Sevilla, Spain}
 
\begin{abstract}
    The elastic and inelastic scattering of $^7$Be from $^{12}$C have been measured at an incident energy of 35 MeV. The inelastic scattering leading to the 4.439 MeV excited state of $^{12}$C has been measured for the first time. The experimental data cover an angular range of $\theta_{cm}$ = 15$^{\circ}$-120$^{\circ}$. Optical model analyses were carried out with Woods-Saxon and double-folding potential using the density dependent M3Y (DDM3Y) effective interaction. The microscopic analysis of the elastic data indicates breakup channel coupling effect. A coupled-channel analysis of the inelastic scattering, based on collective form factors, show that mutual excitation of both $^7$Be and $^{12}$C is significantly smaller than the single excitation of $^{12}$C. The larger deformation length obtained from the DWBA analysis could be explained by including the excitation of $^7$Be in a coupled-channel analysis. The breakup cross section of $^7$Be is estimated to be less than 10$\%$ of the reaction cross section. The intrinsic deformation length obtained for the $^{12}$C$^*$ (4.439 MeV) state is $\delta _2$ = 1.37 fm. The total reaction cross section deduced from the analysis agrees very well with Wong's calculations for similar weakly bound light nuclei on $^{12}$C target. 
\end{abstract}

\date{\today}

\maketitle

\section{Introduction}
The elastic scattering of light stable and unstable nuclei at low energies (near or above the Coulomb barrier) is an important tool to study the change in properties of nuclei as we move towards the drip lines~\cite{BA11,KO16}. The breakup and transfer reaction channels involving such nuclei significantly impact the elastic scattering. This is even more evident, when there is low binding and cluster structure of such nuclei. The loosely bound stable Lithium isotopes $^6$Li and $^7$Li have cluster structures and are well known for their breakup channels~\cite{BA86,SH81,GU99,GU01}. Their breakup thresholds are 1.474 and 2.467 MeV respectively~\cite{SE88}. The $^7$Be nucleus has a similar cluster structure and a breakup threshold of 1.59 MeV. It is radioactive with a half-life of 53.22 days~\cite{TI02}. One can thus expect similar large breakup cross-sections in case of $^7$Be. However, previous elastic scattering studies with  $^7$Be seem to indicate weak breakup couplings at low energy ($<$ 10 MeV/u) and the breakup threshold anomaly does not apply for reactions with $^7$Be~\cite{KO16}. The $^7$Be nucleus behaves more like $^6$Li in contrast to $^7$Li. In addition, for reactions with a light target like $^{12}$C, contribution of nuclear breakup is expected to be more important than Coulomb breakup. The work of Amro \textit{et al.}~\cite{AM07} involving $^7$Be + $^{12}$C at 34 MeV show that the $^7$Be-induced $\alpha$-transfer reactions are more predominant as compared to breakup. Low breakup cross section was also indicated for a $^7$Be + $^{58}$Ni measurement at 21.5 MeV~\cite{MA15}. Since large breakup yield is absent, such reactions can be used as important tools to study high-excitation $\alpha$-clustering states in the residual nuclei in the context of helium-burning process in nuclear astrophysics~\cite{AM07}. The starting point of such a study is the availability of a reliable set of Optical Model Potential (OMP) parameters. In the $^7$Be + $^{12}$C work~\cite{AM07}, the authors recognized the uncertainty in the OMP due to the limited scattering data and therefore they planned further experiments to extract a dedicated set of reliable parameters.

The present work includes the measurement of the elastic scattering angular distribution of $^7$Be + $^{12}$C at 35 MeV over a large angular range. The OMP deduced from the data would be very useful for the analysis of existing and future low energy experiments involving both $^7$Be and $^8$Be. The limitations of the earlier transfer reaction analysis~\cite{AM07} were mainly due to the unavailability of such elastic scattering data. In addition, the inelastic scattering to the 4.439 MeV excited state of $^{12}$C is measured for the first time and analyzed in the distorted-wave Born approximation (DWBA) and coupled-channel (CC) formalisms. Both phenomenological and microscopic analysis in a double-folding model have been carried out and the effects of breakup and other channels on the elastic scattering are also studied.

The organization of the paper is as follows. In Sec II, the experimental details are described. The results part in Sec. III includes the phenomenological and microscopic analysis of the elastic and inelastic data in DWBA and CC formalism. The framework for the microscopic analysis using a DDM3Y interaction is also described. At the end, the total reaction cross section is deduced. In Sec. IV, conclusions and outlook of the present work are discussed. 

\section{Experiment}

The experiment was carried out at the CERN HIE-ISOLDE~\cite{HIE} radioactive ion beam facility, utilizing the Scattering Experiments Chamber (SEC)~\cite{SEC} at the XT03 beamline. The experiment took place at the end of the running period of CERN. A uranium carbide target~\cite{ST19} was irradiated with 0.37$\mu$A of 1.4 GeV protons from the PS-booster offline for 3 days. The activated target was then mounted on the GPS target station and heated up during the present experiment. The $^7$Be was extracted using the RILIS~\cite{RILIS} laser ion-source, and accelerated using the HIE-ISOLDE post accelerator. The stable beam contaminants produced, when the residual gas in the REX-EBIS charge breeder is ionized~\cite{REX}, are mainly $^{14}$N$^{4+}$ and less abundant $^{21}$Ne$^{6+}$, having the same A/q as $^7$Be$^{2+}$ for which the machine was set up. A stripping foil and a dipole before the experimental station was used to clean the beam by further stripping to $^7$Be$^{4+}$, making possible to separate the different components and uniquely select the $^7$Be beam. The potential isobaric contaminant coming from the target is $^7$Li. Any $^7$Li that could have made it into the charge breeder and been accelerated as $^7$Li$^{2+}$ was fully stripped in the foil and could not have passed the dipole. The 35 MeV $^{7}$Be beam of resolution $\sim$ 168 keV and intensity $\sim$ 5 $\times$ 10$^5$ pps was incident on a 15 $\mu$m thick CD$_2$ target. We also used a CH$_2$ target of thickness 15 $\mu$m and a $^{208}$Pb target of thickness 1 mg/cm$^2$ for background measurements and normalization respectively. 

\begin{figure}[h!]
    \centering 
    \includegraphics[width = 0.5\textwidth]{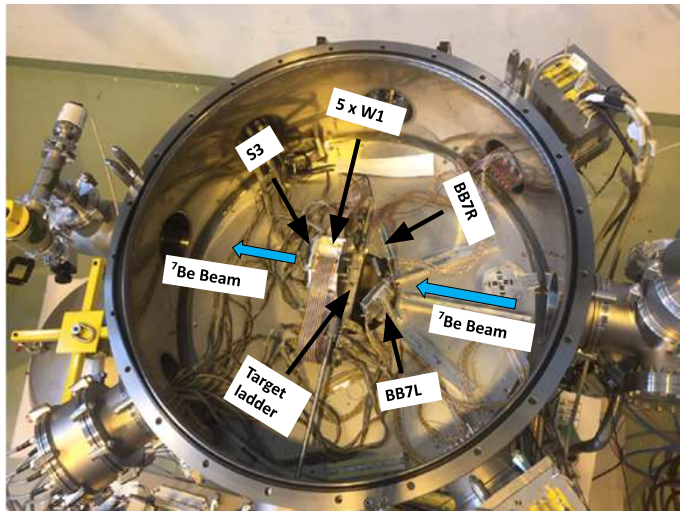}
    \caption{The experimental setup inside the Scattering Experiments Chamber at the XT03 beamline~\cite{SEC}.}
    \label{setup1}
\end{figure}

The charged particles emitted from the reaction were detected by a set of 5 double-sided 16$\times$16 Micron W1 silicon strip detectors (DSSD) of thickness 60 $\mu$m ($\Delta E$) each, backed by MSX25 unsegmented silicon pad detectors of thickness 1500 $\mu$m ($E$). These detectors were placed in a pentagon geometry covering $\theta = 40^{\circ} - 80^{\circ}$ in lab for charged particle detection. The forward angles from 8$^{\circ}$ $-$ 25$^{\circ}$ were covered by a 24 rings $\times$ 32 sectors, Micron S3 annular detector of thickness 1000 $\mu$m. The back angles from 127$^{\circ}$ $-$ 165$^{\circ}$ were covered by two Micron BB7 32$\times$32 DSSDs of thickness 60 $\mu$m and 140 $\mu$m, backed by MSX40 unsegmented silicon pad detectors of thickness 1500 $\mu$m. The experimental setup inside the SEC is shown in Fig. \ref{setup1}. The measurements were made at angular intervals of about 1$^{\circ}$ in S3 range and 2$^{\circ}$ in the pentagon  DSSD range.

The energy calibration of the detectors was done using a $^{148}$Gd$-^{239}$Pu$-^{241}$Am$-^{244}$Cm mixed $\alpha$ source. The dynamic range in the forward detector S3 was much higher. Thus, for calibration at higher energies, we used the elastic peaks from the Rutherford scattering of $^{7}$Be and $^{12}$C beams at ${5}$ MeV/u and ${5.15}$ MeV/u respectively on a $^{208}$Pb  target. The grazing angle of $^7$Be + $^{208}$Pb at 35 MeV is around 100$^{\circ}$. Thus, the experimental data in one of the forward rings of the annular detector were normalized to Rutherford cross sections with $^{208}$Pb, and that ring was used as a monitor detector. We took very short consecutive runs of $^7$Be on $^{208}$Pb and CD$_2$ targets where the fluctuation in beam intensity was within 10$\%$. In addition to the statistical uncertainty in the yields, the present data include 10$\%$ error due to beam intensity and 10$\%$ error due to target thickness. 

\begin{figure}[h!]
    \centering 
    \includegraphics[width = 0.5\textwidth]{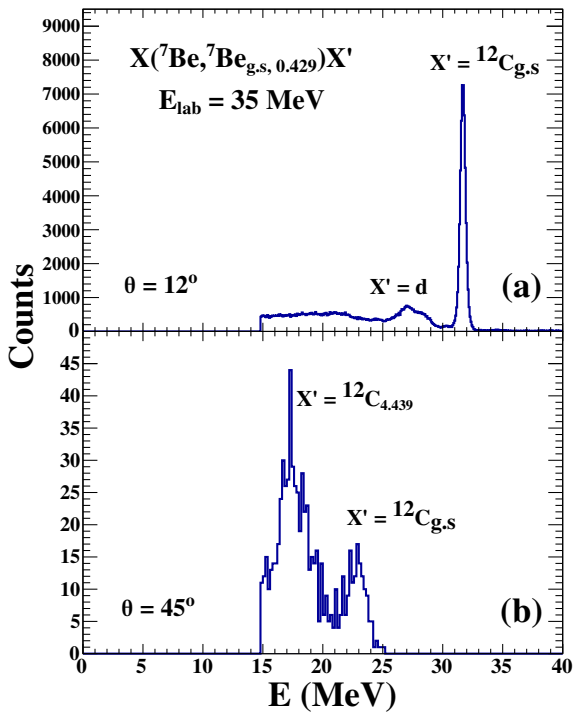} 
    \caption{Energy spectrum for elastic and inelastic scattering of $^{7}$Be + $^{12}$C at 35 MeV. Panel (a) shows elastic $^{7}$Be scattered from $^{12}$C and d in one of the rings of S3, at $\theta = 12^{\circ}$. Panel (b) shows elastic as well as inelastic peak corresponding to 4.439 MeV state of $^{12}$C in one pentagon DSSD, at $\theta = 45^{\circ}$. The bound excited state of $^7$Be at E$_x = 0.429$ MeV (1/2$^-$) could not be separated from the ground state, implying the data to be quasi-elastic.}
    \label{E7BeS3}
\end{figure}

\section{Results}

The elastic peak from $^7$Be + $^{12}$C scattering is distinctly visible in each ring of the forward annular detector. Fig. \ref{E7BeS3}(a) shows the energy spectrum at $\theta = 12^{\circ}$, delineating the prominent elastic peak. The elastic $^7$Be scattered from deuterons in the CD$_2$ target is also seen as a small broad peak in the spectrum. In the pentagon DSSDs, the scattered $^7$Be deposits all its energy in the $\Delta$E detectors of the telescopes. Fig.~\ref{E7BeS3}(b) is the energy spectrum at $\theta = 45^{\circ}$ showing the $^7$Be elastic peak as well as the inelastic scattering to the 4.439 MeV state of $^{12}$C. The ground state peak is separated from the 2$^+$ inelastic peak as seen in Fig. \ref{E7BeS3}(b). However, the bound excited state of $^7$Be at E$_x = 0.429$ MeV (1/2$^-$) could not be separated from the ground state. Thus the present data correspond to quasi-elastic scattering of $^7$Be + $^{12}$C at 35 MeV. In this context, it may be noted that the inelastic scattering to the 0.429 MeV state of $^7$Be is expected to be negligible in comparison to the $^7$Be + $^{12}$C elastic scattering~\cite{BA11,PE95}. In the forward angles 8$^{\circ}$ $-$ 25$^{\circ}$, due to several overlapping channels and non-availability of particle identification, inelastic scattering cross sections to the 4.439 MeV state of $^{12}$C could not be extracted. At the back angles 127$^{\circ}$ $-$ 165$^{\circ}$, the elastic and inelastic data could not be separated due to small differences in energies as well as overlap with the (d,p) data. Hence, these data are not included in the present work.  
 
\begin{figure}[h!]
    \centering 
        \includegraphics[width = 0.5\textwidth]{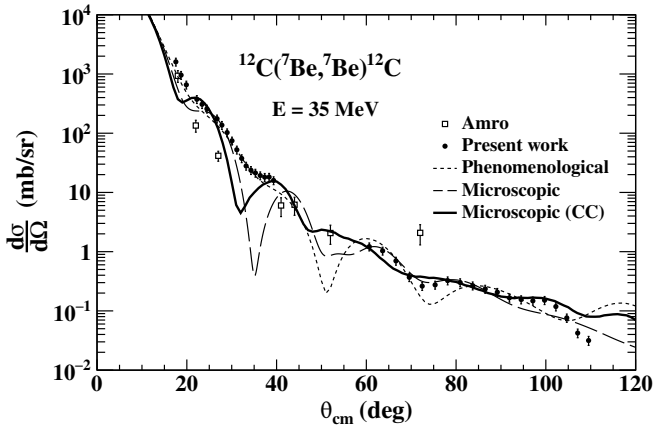}\\[\abovecaptionskip]
        \caption{Elastic scattering angular distribution of $^{7}$Be + $^{12}$C at E = 35 MeV. The  microscopic (phenomenological) fits to the data are given by the dashed (dotted) curves. The coupled-channel calculations are given by the solid curve. The data of Amro {\it et al.}~\cite{AM07} are also shown for comparison. The bound excited state of $^7$Be at 0.429 MeV could not be separated from the ground state, implying the data to be quasi-elastic.}
        \label{Elastic}
\end{figure}

The angular distribution of $^7$Be + $^{12}$C elastic scattering at 35 MeV is shown in Fig. \ref{Elastic}. For comparison, the data of Amro \textit{et al.}~\cite{AM07} at 34 MeV is also plotted in the figure. If we compare our data of much better energy and angular resolution,  with Amro's data at the forward angles, we observe that the elastic data at $\sim$ 18$^{\circ}$ agree with our data within error bars but for the other two angles, it is somewhat lower in value. At around 72$^{\circ}$, Amro's data gives a higher value of cross section than our data. This discrepancy can be attributed to uncertainties in data normalization arising from the large beam energy spread ($\sim$ 1 MeV) and, more importantly to the angular divergence at target position ($\sim$ 6$^{\circ}$) in~\cite{AM07}. Their limited angular range of scattering data resulted in difficulty to extract reliable OMP for subsequent study of the transfer reactions. Analysis of the $\alpha$- and n-transfer reactions from the $^{12}$C and d in the CD$_2$ target from the present experiment will be presented elsewhere.

Optical model analyses of the data were carried out with the code FRESCO~\cite{TH88}. First, the data were analyzed using volume type Woods-Saxon potentials and Coulomb potential due to uniformly charged spheres ($r_C = 1.25$ fm). The potential parameters are given in Table \ref{tab:OMP}. The starting point of the phenomenological analysis was the OMP parameters from $^7$Li + $^{12}$C elastic scattering at 34 MeV~\cite{CO86}. The dotted curve in Fig. 3 corresponds to the phenomenological fit to the present data. The phenomenological fit agrees satisfactorily with the data in the whole angular range, except around 75$^{\circ}$ and angles greater than 100$^{\circ}$. The fit also agrees with the data of Amro \textit{et al.}~\cite{AM07} at 18$^{\circ}$ and 45$^{\circ}$ and explains overall the nature of their data except at 72$^{\circ}$. It is to be noted that their data point at 72$^{\circ}$ also did not agree with their optical model calculations~\cite{AM07}. 

To study the data further and have an estimate of the breakup effects on the elastic scattering, we carried out a microscopic analysis of the data. Here, the two-body potential $V(r)$ is generated  in a double folding model using densities of the $^{7}$Be and $^{12}$C along with the density dependent M3Y (DDM3Y) effective interaction~\cite{DU14,BA03}. The densities of $^{7}$Be and $^{12}$C used in the present work are obtained from variational Monte Carlo calculations using the Argonne v18 two-nucleon and Urbana X three-nucleon potentials (AV18+UX)~\cite{WI91}. Earlier, the DDM3Y effective interaction has been successfully used to describe nuclear matter~\cite{BA04}, radioactivity~\cite{BA05}, proton scattering~\cite{GU06} as well as resonances in unstable~\cite{DU14} and unbound nuclei~\cite{DU18}.\\

\begin{table*}[h!]
    \caption{Optical model potential parameters for elastic scattering of $^7$Be + $^{12}$C at E = 35 MeV, $r_C = 1.25$ fm;\\ $R_x = r_x (A_P^{\frac{1}{3}} + A_T^{\frac{1}{3}})$}
    \label{tab:OMP}
    \begin{ruledtabular}
        \begin{tabular}{c  c  c  c  c  c  c  c  c  c  c}
            $E$ &$V_{r}$ &$r_{r}$ &$a_{r}$ &$W_{V}$ &$r_{i}$ &$a_{i}$ & $\chi^2 _{el}/N$\\
            (MeV) &  (MeV)  &  (fm)   &  (fm)   &  (MeV)  &  (fm)   &  (fm)   &  & \\
            \hline \\
             35   & 142.96  &  0.702  &  0.617  &  9.95  &  1.041  &  1.079  &  2.6 \\
        \end{tabular}
    \end{ruledtabular}
\end{table*}

\begin{table*}[h!]
    \caption{Renormalization factors ($N_{R}$, $N_{I}$) of DDM3Y folded potential for $^7$Be + $^{12}$C scattering at E = 35 MeV. Excited state energy ($E_x$), angular momentum transfer ($l$),  deformation length ($\delta _2$), deformation parameter ($\beta _2$) and $\chi^2$/N from best fits to the elastic and inelastic scattering data. $\delta _2 = \beta _2 R \mbox{, where } R = 1.2 \times 12^{\frac{1}{3}}$}.
    \label{tab:micro}
    \begin{ruledtabular}
        \begin{tabular}{c  c  c  c  c  c  c  c  c  c  c  c  c }
            Nucleus & Calculation & & $E$   & $E_x$  & $N_{R}$ & $N_{I}$ & $l$ & $\delta _2$ & $\beta _2$ & $\chi^2 _{el}/N$ & $\chi^2 _{inel}/N$\\
                    &    &    & (MeV) & (MeV) &     &     &     & (fm) &     &   &\\
            \hline \\
            $^{12}$C & DWBA   &   & 35   & 4.439  &  0.68  &  0.30  &  2  & 1.83 & 0.67 &  9.3  & 4.5 \\
            \hline \\
            $^{12}$C & CC   &   & 35   & 4.439  &  1.00  &  0.20  &  2  & 1.37 & 0.49 &  8.0  & 4.8 \\
        \end{tabular}
    \end{ruledtabular}
\end{table*}

The double-folded potential between two nuclei is given by 

\begin{equation}
   V(R) = \int{\rho_1(\boldsymbol{r_1})\rho_2(\boldsymbol{r_2})v(\boldsymbol{s},\rho_1,\rho_2,E)d^3r_1d^3r_2}
\end{equation}
where $\rho_i$, $i=1,2,$ is the nucleon point-density distribution for the projectile and the target, $R$ is the separation between them, $\boldsymbol{s}=\boldsymbol{R}-\boldsymbol{r_1} - \boldsymbol{r_2}$, and $v$ is the density- and energy-dependent effective nucleon-nucleon interaction. 
Assuming that the density and the radial dependence of $v$ can be separated, one can write~\cite{BA03}
\begin{equation}
    v(s,\rho_1,\rho_2,E) = t^{M3Y}(s,E)g(\rho_1,\rho_2,E)
\end{equation}
The density-independent part $t^{M3Y}(s, E)$ contains the M3Y interaction and the zero-range pseudo potential,
\begin{equation}
    t^{M3Y}(s,E)=7999\frac{e^{-4s}}{4s} - 2134\frac{e^{-2.5s}}{2.5s} + J_{00}(E)\delta(s)
\end{equation}
with
\begin{equation}
    J_{00}(E)=-276(1-0.005 E/A) 
\end{equation}
The density-dependent part $g(\rho_1,\rho_2,E)$ is as follows, 
\begin{equation}
     g(\rho_1,\rho_2,E) = C(1-b(E)\rho_{1}^{2/3})(1-b(E)\rho_{2}^{2/3})
\end{equation}
Here $C$ has been taken as 1.0.  The other constant of the interaction $b$ when used in single folding model description, is determined by nuclear matter calculations~\cite{BA04} as 1.624 fm$^2$. This constant has also been used in scattering~\cite{GU06} and bound state studies~\cite{DU14,DU18}.

\begin{figure}[h!]
    \centering 
        \includegraphics[width = 0.5\textwidth]{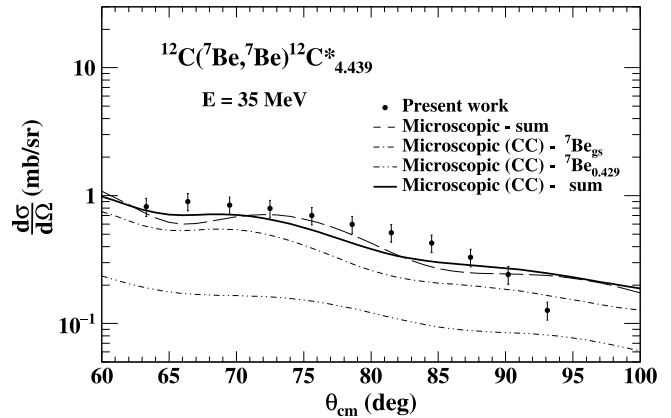}\\[\abovecaptionskip]
    \caption{Inelastic scattering angular distribution of $^{7}$Be + $^{12}$C to the 4.439 MeV (2$^+$) state at E = 35 MeV. The DWBA calculations using microscopic potentials are given by the dashed curve and includes the sum of single and mutual excitations. The CC calculations are given by the dash-dot (dash-dot-dotted) curves for single (mutual) excitations respectively. The sum of the two contributions are shown by the solid curve.}
        \label{Inelastic}
\end{figure}

In the present microscopic analysis, both the real ($V$) and volume imaginary ($W$) parts of the folded nuclear potentials are assumed to have the same shape, i.e. $V_{\rm micro}(r)~=~V + iW$ = ($N_{\rm R}$ + $iN_{\rm I}$)$V$($R$) where, $N_{\rm R}$ and $N_{\rm I}$ are the renormalization factors for real and imaginary parts respectively, obtained by fitting the elastic scattering data. The microscopic fit using FRESCO is shown by the dashed curve in Fig. \ref{Elastic}. A satisfactory fit is obtained for an $N_{\rm R}$ value of 0.68 and an $N_{\rm I}$ value of 0.30 (Table \ref{tab:micro}). The $N_{\rm R}$ value of less than unity suggests coupling of the breakup channel to elastic scattering~\cite{SA79,SA83}. The $N_{\rm I}$ value is attributed to the absorption of flux due to inelastic excitation, breakup or other reaction channels. The microscopic fit shows a more diffractive elastic angular distribution than observed, particularly at $\sim$ 35$^{\circ}$. However, it provides a better description of the data at larger angles compared to the phenomenological fit. The microscopic calculation also explains the nature of the data of Amro {\it et al.}~\cite{AM07} better than the phenomenological calculation. 

To better understand the effects of nonelastic channels, coupled-channel (CC) calculations were also carried out. In addition to the ground state of the projectile and target nuclei, some low-lying excited states of these nuclei were included in the calculations. In particular, the modelspace included the $^{7}$Be($1/2^-$) bound excited state at $E_x=0.429$~MeV, the narrow $7/2^-$ low-lying $^{7}$Be resonance at $E_x=4.57$~MeV and the $2^{+}$ state at $E_x= 4.439$~MeV of $^{12}$C. It is worth noting that the inclusion of the inelastic couplings will also affect the elastic scattering cross section. In $\chi^2$-fitting, the $N_R$ and $N_I$ of the bare projectile-target potential were taken as adjustable parameters to best reproduce the measured elastic scattering data. This resulted in $N_R=1$ and $N_I=0.2$ (Table \ref{tab:micro}). The reduced value of $N_I$ with respect to that obtained above is consistent with the explicit inclusion of channels removing flux from the elastic channel. The dip around 30$^{\circ}$ is much reduced in CC calculation and is shown by the solid curve in  Fig. \ref{Elastic}. 

The inelastic scattering angular distribution of $^7$Be + $^{12}$C to the 4.439 MeV (2$^+$) state, measured for the first time, is shown in Fig. \ref{Inelastic}. It may be noted that, in this angular range, the inelastic cross section is of comparable magnitude to the elastic cross section. Since the grazing angle is $\sim$ 8$^{\circ}$, there is negligible contribution of Coulomb excitation over the angular range of the inelastic scattering data. The nucleus-nucleus optical potentials obtained from fitting the elastic scattering data were used to generate the distorted waves for the analysis of inelastic scattering in DWBA formalism using FRESCO~\cite{TH88}. The experimental resolution of the present work did not allow the separation of the $^{7}$Be$_\text{g.s.}$+$^{12}$C$_{4.439}$ channel from the mutual excitation  $^{7}$Be$_{0.429}$+$^{12}$C$_{4.439}$ channel. Thus, to include the $^{7}$Be$_{0.429}$ excitation in the DWBA analysis, contribution from these two channels were calculated and added together (sum), given by the dashed curve in Fig. \ref{Inelastic}. The collective rotor model is used to obtain the transition form factors. The deformation length $\delta_2$ is obtained by fitting the inelastic scattering angular distributions. The microscopic analysis could explain the inelastic data very well, while the phenomenological analysis does not produce a satisfactory fit.  Table \ref{tab:micro} gives the excited state energy ($E_x$), angular momentum transfer ($l$), renormalization factors ($N_R, N_I$), deformation length ($\delta_2$), deformation parameter ($\beta_2$), and $\chi^2$/N for the analyses. From DWBA analysis using microscopic potentials, the $\delta_2$ obtained is 1.83 fm, as compared to the reported values of $\sim$ 1.4 fm~\cite{CO86,ZE80}.  

To explain the larger deformation length, we extended the CC calculations including the excitation of the $^7$Be nucleus. As mentioned earlier, the three included states of $^{7}$Be are considered members of a $K=1/2^-$ rotational band. They are coupled together by deforming the projectile-target microscopic folding potential using a quadrupole deformation length $\delta_2=2$~fm adopted from~\cite{KI20}. Likewise, the ground and first $2^+$ state of $^{12}$C are considered members of a $K=0$ rotational band and coupled together by deforming the projectile-target potential with a deformation length of $\delta_2$, obtained from best fit to the present data. Couplings between the included states were considered to all orders, thus performing a full coupled-channels calculation. 
The CC calculations are compared to the inelastic data in Fig. \ref{Inelastic}. For a meaningful comparison, the single and mutual excitations have been considered in the calculations and added together (sum), shown by the solid curve. It becomes apparent from Fig. \ref{Inelastic} that the mutual excitation (dash-dot-dotted curve) is significantly smaller than the single excitation (dash-dot curve) but its inclusion improves the agreement with the data. There is slight underestimation, which might be due to the simplified rotor model assumed for the $^{7}$Be and $^{12}$C nuclei. From the best fit to the present data, we get $\delta_2$ = 1.37 fm (Table \ref{tab:micro}). A 10$\%$ variation of $\chi^2_{inel}$/N results in a variation of 0.13 fm in the value of $\delta_2$ in agreement to the existing values~\cite{CO86, ZE80}. This results in a deformation parameter $\beta_2$ of 0.49 with a variation of 0.05. 
 
In an alternative approach, one may treat the reaction in an extended three-body model, comprising of the two-body projectile ($\alpha$+$^{3}$He) plus the target, similar to that used in the continuum discretized coupled-channels (CDCC) calculations~\cite{HA22}. In this model, target excitations are included by means of deformed $^3$He+$^{12}$C and $^4$He+$^{12}$C potentials. In this way, the projectile excitation (and breakup) would be treated within the cluster model whereas target excitations would be treated in the collective model. These extended CDCC calculations were initially proposed and implemented in~\cite{YA86} and have been recently revisited and applied to several systems~\cite{CH15,GO17}. The application to the present reaction is ongoing and the results will be presented elsewhere. 

\begin{figure}[h!]
    \centering 
    \includegraphics[width = 0.5\textwidth]{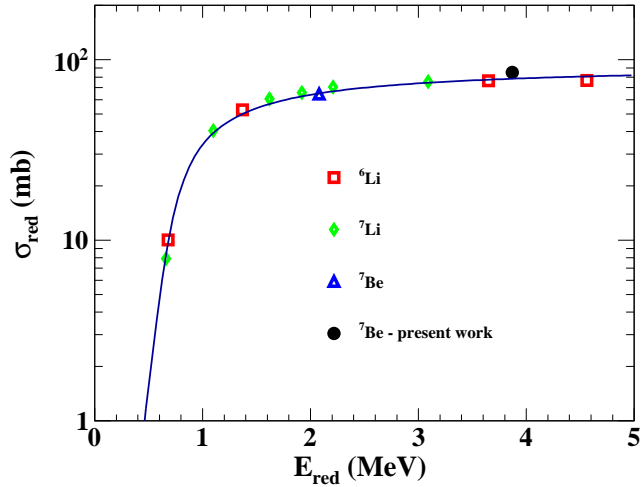}
    \caption{The reduced reaction cross sections for the $^{7}$Be + $^{12}$C system obtained in the present work along with reduced reaction cross sections of the lithium isotopes on $^{12}$C. The solid curve represents Wong's expression in Eqn. \ref{eq:wong}.}
    \label{Red_tcsec}
\end{figure}

To investigate further the dynamics of $^{7}$Be, we deduced the total reaction cross section from the elastic scattering analysis. As suggested by Kolata~\cite{KO09} and Aguilera~\textit{et al.}~\cite{AG11}, the reduced total reaction cross section of light stable and unstable nuclei on $^{12}$C can be compared with that given by Wong~\cite{WO73},

\begin{equation}
    \sigma_{red} = \frac{\epsilon_{0}r_{0}^{2}}{2E_{red}} \ln \Bigg[ 1+ \exp\Bigg(\frac{2\pi (E_{red}-V_{red})}{\epsilon_{0}}\Bigg) \Bigg]
    \label{eq:wong}
\end{equation}

where the cross section is in fm$^2$, $\epsilon_0 = \hbar \omega_0 \frac{(A^\frac{1}{3}_{P} + A^\frac{1}{3}_{T})}{Z_P Z_T}$ and $V_{red} = V_0 \frac{(A^\frac{1}{3}_{P} + A^\frac{1}{3}_{T})}{Z_P Z_T}$ are denoted as the Wong-model parameters and the reduced energy is given by
\begin{equation}
    E_{red} = E_{cm}\frac{(A_P^{\frac{1}{3}} + A_T^{\frac{1}{3}})}{Z_P Z_T}. 
\end{equation}
These parameters were obtained in~\cite{BA11} by fitting the data for $^{6}$He, $^{6}$Li, $^{7}$Li, $^{7}$Be, $^{8}$Li, $^{8}$B, $^{9}$Be and $^{11}$Be. The parameters were found to be $V_{red} = 0.64(3)$ MeV, $r_0 = 1.73(2)$ fm and $\epsilon_0 = 0.43$ MeV. In particular, the value $r_0= 1.73(2)$ fm is significantly larger than the typical values of $1.4 - 1.5$ fm. The reduced total reaction cross section for $^{6,7}$Li and $^7$Be + $^{12}$C obtained  in~\cite{BA11}, along with the fitted curve is plotted in Fig. \ref{Red_tcsec}. The calculated $^{7}$Be total reaction cross section at 35 MeV for CC calculations using microscopic potentials is 1503 mb and the reduced total reaction cross section is 85 mb. The calculated value agrees very well with Wong's calculation on the existing $^{6,7}$Li and $^7$Be data. Preliminary CDCC calculations show that for $^7$Be, the breakup and inelastic cross section to $^7$Be bound excited state are respectively 9.12$\%$ and 0.67$\%$ of the reaction cross section. The absorption cross section is 90.2$\%$ of the reaction cross section.

\section{Conclusion}
The elastic and inelastic scattering of $^7$Be + $^{12}$C at 35 MeV is studied in the present work. The experimental data as well as the deduced optical model potential parameters can be useful for studies on transfer reactions involving $^7$Be and $^8$Be at these energies, particularly in relation to nuclear astrophysics. This work also reports the first inelastic scattering measurement of $^7$Be + $^{12}$C to the 4.439 MeV excited state of $^{12}$C. To have an estimate of the influence of other channels like breakup on the elastic scattering, we also carried out a coupled channel analysis of the data. The microscopic calculation for the elastic scattering gives better fit, particularly at larger angles as compared to the phenomenological fit with Wood-Saxon potentials. The $N_R$ value of less than unity in the microscopic fit indicates the presence of breakup channel coupling effects. Coupled-channel calculations improve on the fits and the value of $N_R$ becomes compatible with 1. For inelastic scattering, mutual excitation of both $^7$Be and $^{12}$C is significantly smaller than the single excitation of $^{12}$C but its inclusion improves the agreement of the calculations with the inelastic data. The larger deformation length obtained from the DWBA analysis could be explained by including the excitation of $^7$Be in a coupled-channel analysis. The breakup cross section of $^7$Be is estimated to be less than 10$\%$ of the reaction cross section. The reduced total reaction cross section from the present work, agrees very well with Wong's calculation on existing $^{6,7}$Li and $^7$Be scattering data. The present study is very important to further understand the $\alpha$-cluster transfer reactions in nuclear astrophysics. \\

\noindent
{\large{\bf Acknowledgement}}\\

The authors thank the ISOLDE engineers in charge, RILIS team and Target Group at CERN for their support. D. Gupta acknowledges research funding from the European Union's Horizon 2020 research and innovation programme under grant agreement no. 654002 (ENSAR2) and ISRO, Government of India under grant no.
ISRO/RES/2/378/15$-$16. O. Tengblad would like to acknowledge the support by the Spanish Funding Agency (AEI / FEDER, EU) under the project PID2019-104390GB-I00. 
I. Martel would like to acknowledge the support by the Ministry of Science, Innovation and Universities of Spain (Grant No. PGC2018-095640-B-I00). 
 J. Cederkall acknowledges grants from the Swedish Research Council (VR) under contract numbers VR-2017-00637  and  VR-2017-03986 as well as grants from the Royal Physiographical Society.
J. Park would like to acknowledge the support by Institute for Basic Science (IBS-R031-D1). 
S. Szwec acknowledges support by the Academy of Finland (Grant No. 307685). 
A.M.M. is supported by the  I+D+i project PID2020-114687GB-I00 funded by MCIN/AEI/10.13039/501100011033, by the grant Group FQM-160 and by project P20\_01247, funded by the Consejer\'{\i}a de Econom\'{\i}a, Conocimiento, Empresas y Universidad, Junta de Andaluc\'{\i}a (Spain) and by ``ERDF A way of making Europe".


\begin{thebibliography}{50}

\bibitem{BA11} A. Barioni \textit{et al.}, Phys. Rev. C 84, 014603 (2011); see references therein.
\bibitem{KO16} J. J. Kolata \textit{et al.}, Eur. Phys. J. A 52, 123 (2016).
\bibitem{BA86} G. Baur \textit{et al.}, Nuclear Physics A 458 188-204 (1986).
\bibitem{SH81} A. C. Shotter \textit{et al.}, Phys. Rev. Lett. 46 12 (1981).
\bibitem{GU99} D. Gupta \textit{et al.}, Nuclear Physics A 646 161-174 (1999).
\bibitem{GU01} D. Gupta \textit{et al.}, Nuclear Physics A 683 1-4 (2001).
\bibitem{SE88} A. J. Selove, Nuclear Physics A 490 1-225 (1988).
\bibitem{TI02} D. R. Tilley {\it et al.}, Nucl. Phys. A 708 3-163 (2002).
\bibitem{AM07} H. Amro \textit{et al.}, Eur. Phys. J. Special Topics 150, 1-4 (2007).
\bibitem{MA15} M. Mazzocco \textit{et al.}, Phys. Rev. C 92, 024615 (2015).
\bibitem{HIE} https://hie-isolde-project.web.cern.ch/hie-isolde-project/
\bibitem{SEC}  https://isolde.cern/sec
\bibitem{ST19} Thierry Stora, private communication, CERN EDMS MEDICIS operation report 2093202 (MED-OP-18-019).
\bibitem{RILIS} V. Fedosseev \textit{et al.}, J. Phys. G: Nucl. Part. Phys. 44 (2017) 084006 (28pp)
\bibitem{REX} F. Wenander \textit{et al.}, Review of Scientific Instruments 77, 03B104 (2006).
\bibitem{PE95} I. Pecina \textit{et al.}, Phys. Rev. C 52, 191 (1995).
\bibitem{TH88} I. J. Thomspon, Comput. Phys. Rep. 7, 167 (1988).
\bibitem{CO86} J. Cook \textit{et al.}, Phys. Rev. C 33, 915 (1986).
\bibitem{WI91} R. B. Wiringa, Phys. Rev. C 43, 1585 (1991); https://www.phy.anl.gov/theory/research/density/
\bibitem{BA04} D. N. Basu, J. Phys. G: Nucl. Part. Phys. 30 B7 (2004).
\bibitem{BA05} D. N. Basu, P. Roy Chowdhury and C. Samanta Phys. Rev. C 72 051601 (2005).
\bibitem{GU06} D. Gupta, E. Khan and Y. Blumenfeld, Nucl. Phys. A 773, 230 (2006).
\bibitem{DU18} S. K. Dutta, D. Gupta, Swapan K Saha, Phys. Lett. B 776, 464 (2018); see references therein.
\bibitem{DU14} S. K. Dutta, D. Gupta, D. Das, Swapan K Saha, Jour. Phys. G: Nucl. Part. Phys. 41, 095104 (2014); see references therein.
\bibitem{BA03} D. N. Basu, Phys. Lett. B 566, 90 (2003).
\bibitem{SA79} G. R. Satchler, W.G. Love, Phys. rep. 55 (1979) 183.
\bibitem{SA83} Y. Sakuragi \textit{et al.}, Prog. Theor. Phys. 70 (1983) 1047; Prog. Theor. Phys. 68 (1982) 322.
\bibitem{ZE80} A. F. Zeller \textit{et al.}, Phys. Rev. C 22, 1534 (1980); see references therein.

\bibitem{KI20} G. G. Kiss \textit{et al.}, Phys. Lett. B 807, 135606 (2020). 
\bibitem{HA22} K. Hagino \textit{et al.}, Progress in Particle and Nuclear Physics 125, 103951 (2022); see references therein.
\bibitem{YA86} M. Yahiro \textit{et al.}, Prog. Theor. Phys. Suppl. 89, 32 (1986).
\bibitem{CH15} H.-T. Pierre Chau, Eur. Phys. J. A 51, 1 (2015).
\bibitem{GO17} M. G\'omez-Ramos and A. M. Moro, Phys. Rev. C95, 034609 (2017)
\bibitem{KO09} J. J. Kolata and E. F. Aguilera, Phys. Rev. C 79, 027603 (2009).
\bibitem{AG11} E. F. Aguilera, \textit{et al.}, Phys. Rev. C 83, 021601 (2011).
\bibitem{WO73} C. Y. Wong, Phys. Rev. Lett. 31, 766 (1973).

\end{thebibliography}
\end{document}